\documentclass[aps,prl,twocolumn,showpacs,superscriptaddress,groupedaddress]{revtex4-1}  
\usepackage{graphicx}  
\usepackage{dcolumn}   
\usepackage{bm}        
\usepackage{amssymb}   
\usepackage{txfonts}
\usepackage{upgreek}
\usepackage{CJK,graphicx}
\usepackage[colorlinks=true,urlcolor=blue,linkcolor=blue,citecolor=blue,dvipdfm]{hyperref}%
\usepackage[all]{hypcap}

\begin{document}


\title{Borophane: Stable Two-dimensional Anisotropic Dirac Material with Ultrahigh Fermi Velocity}

\author{Li-Chun Xu}
 \affiliation{
College of Physics and Optoelectronics, Taiyuan University of Technology, Taiyuan 030024, China
}
\author{Aijun Du}
 \email{aijun.du@qut.edu.au}
\author{Liangzhi Kou}
 \email{liangzhi.kou@qut.edu.au}
\affiliation{
School of Chemistry, Physics, and Mechanical Engineering Faculty, Queensland University of Technology,
Garden Point Campus, QLD 4001, Brisbane, Australia
}

\date{\today}

\begin{abstract}
Recent synthesis of monolayer borophene (triangle boron monolayer) on the substrate opens the era of boron nanosheet (Science, 350, 1513, $\mathbf{2015}$), but the structural stability and novel physical properties are still open issues. Here we demonstrated borophene can be stabilized with fully surface hydrogenation, called as borophane, from first-principles calculations. Most interesting, it shows that borophane has direction-dependent Dirac cones, which are mainly contributed by in-plane \emph{p$_{x}$} and \emph{p$_{y}$} orbitals of boron atoms. The Dirac fermions possess an ultrahigh Fermi velocity up to 3.0$\times$10$^{6}$ m/s, 4 times higher than that of graphene. The Young's modules are calculated to be 129 and 200 GPa$\cdot$nm along two different directions, which is comparable with steel. The ultrahigh Fermi velocity and high mechanical feature render borophane ideal for nanoelectronics applications.
\end{abstract}

\maketitle

The Dirac material is a special kind of condense matter, which is characterized with linear energy dispersion at Fermi level. The electrons in Dirac materials display a linear energy-momentum relativistic dispersion accurately described by the massless Dirac Hamiltonian. This unique electronic structure results in many exceptional physical properties\cite{novoselov2005two,wehling2014dirac}, like exceedingly large charge-carrier mobility massless fermions\cite{bolotin2008ultrahigh}, quantum Hall effects\cite{zhang2005experimental} and many other properties\cite{zhang2009topological}, which offers a broad perspective for the development of next-generation nanoscale electronic devices. The most representative Dirac material is graphene, where the Dirac cone is originated from the unpaired \emph{p$_{z}$} orbital with Fermi velocity of 8.2$\times$10$^5$ m/s, among the highest record until now. Beside graphene, many two-dimensional (2D) materials have been theoretically confirmed to possess a Dirac cone\cite{kim2012graphyne,mei2012first,malko2012two,zhou2014semimetallic,yin2013r,wang2015phagraphene,li2014gapless,malko2012competition}, like silicene, germanene (graphene like silicon and germanium) and several graphynes (\emph{sp}-\emph{sp$^2$} carbon allotropes)\cite{xu2014two,malko2012competition}. Unfortunately, only the Dirac cone of graphene has been experimentally confirmed.

Boron is an element that sites left beside carbon on the periodic table. Its chemical complexity and three valance electrons result in at least 16 forms of bulk boron allotropes and numerous low dimensional allotropes, which show abundant promising properties\cite{oganov2009ionic}. Although there are many phases of boron bulk achieved experimentally, all 2D boron nanosheets are proposed theoretically and not confirmed from experiments yet. Zhou et al. have theoretically predicted a novel 2D boron allotrope, which was identified to have a distorted Dirac cone\cite{zhou2014semimetallic}. Recently, an atomically thin 2D boron sheets, named as borophene, has been synthesized on Ag (111) surfaces by physical vapor deposition\cite{mannix2015synthesis}, which open a new era to explore its novel mechanical and physical properties. The structure of borophene can be derived from a hexagonal grid of boron atoms, with additional boron atoms located centrally above each B$_{6}$ hexagon. These films also have considerable toughness, the in-plane Young's modulus along \textbf{a} axis can rival those of graphene. Unlike bulk boron allotropes, borophene is revealed to be a highly anisotropic 2D metal, in addition to its extraordinary structural and mechanical properties. With these aspects, borophene layers have the potential to be an important 2D material in future nanodevices. However, unlike graphene, the free standing borophene is dynamically unstable from previous calculation\cite{mannix2015synthesis}. Meanwhile, the chemical active surface of borophene indicates that it is easy to be affected/oxidized by the environments.  These shortcomings have to be addressed before its real applications. One of feasible solutions is surface functionalization\cite{zhang2012first,sachdev2015materials}, which is an effective approach to chemically modify the properties of the materials\cite{zhang2015hydrogenated}, and enhance structural stability.  Upon the adsorption of hydrogen, 2D Dirac materials, graphene, silicene and germanene becomes wide bandgap semiconductors\cite{Elias610,houssa2011electronic} due to removal of unsaturated orbital. Because the B-H group can be considered as the isoelectronicity of C, hydrogenation is highly expected to modify the property of the metallic borophene.

In this paper, the properties of hydrogenated borophene, so called borophane, are investigated using first-principles calculations. Borophane is shown to be dynamically stable, and is identified to have a distorted Dirac cone like graphene. In particular, this 2D material has the ultrahigh Fermi velocities related to the linear bands near the Dirac points, which rival those of graphene. Meanwhile the outstanding mechanical properties of borophane, which is comparable to these of steel, render it great potential applications in future nanodevices.

Structure relaxations and electronic properties calculations were performed using density functional theory (DFT)\cite{Kohn1965} with the projector-augmented-wave (PAW) method\cite{Blochl1994,Kresse1999}, implemented in the Vienna ab initio simulation package (VASP)\cite{Kresse1996,Kresse1996a}. The exchange-correlation energy was treated within the generalized gradient approximation (GGA), using the Perdew, Burke, and Ernzerhof (PBE) functional\cite{pbe1996,pbe1997}. A kinetic cutoff energy 520 eV of the plane-wave basis was adopted. Brillioun zone integrations were carried out by Monkhorst-Pack k-point\cite{mp} mesh (66$\times$45$\times$1), and structure relaxations were performed using the quasi-Newton algorithm until the force on each atom was less than 1$\times$10$^{-6}$ eV/{\AA}. The phonon dispersion relation have been calculated by the phonopy code\cite{phonopy}. A 3$\times$3$\times$1 supercell was used. The force constants of the borophane were generated based on Density functional perturbation theory(DFPT) by the VASP code.

\begin{figure}
\includegraphics{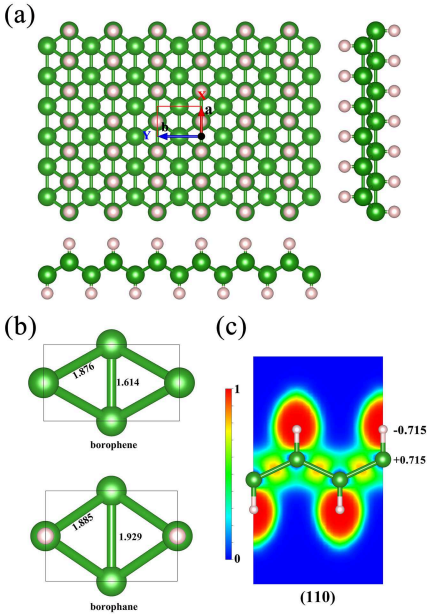}
\caption{\label{fig1}(a)The optimized borophane with top and side views. The unit cell is marked with a red box. The green and white balls represent B and H atoms, respectively. (b)B-B bond length in borophane and borophene unitcells. (c)ELF schematic of borophane (110) surface, where the electron transfer is shown after hydrogenation.}
\end{figure}

Based on the buckled structure of borophene with anisotropic corrugation, where the adjacent row boron atoms are wrinkled alternatively along zigzag direction, we thus passivated each adjacent row along zigzag direction with hydrogen on the opposite sides, rather than neighboring two atoms in graphane.  The fully relaxed borophane was displayed in Fig.~\ref{fig1}.  The unit cell is marked by the red solid rectangular.The optimized lattice constants \textbf{a} and \textbf{b} equal to 1.93 {\AA} and 2.81 {\AA}, respectively. It is noted that the lattice constant \textbf{a} is remarkably increased compared with the values (1.667 {\AA}) in borophene, while the lattice \textbf{b} keeps almost unchanged (2.89 {\AA})\cite{mannix2015synthesis}. As a result, the B-B bond length in borophane is also significantly stretched (0.315{\AA} along \textbf{a} direction, Fig.~\ref{fig1}b), which will affect its mechanical properties as shown in the following. The boron atoms in borophane are constructed from distorted B$_{7}$ clusters. Each B-B bond connects boron atoms with hydrogen attached at the opposite sides of the plane. The buckling height is about 0.8 {\AA}, which agrees well with the buckling height of the borophene on the Ag(111) substrate\cite{mannix2015synthesis}.

\begin{figure}
\includegraphics{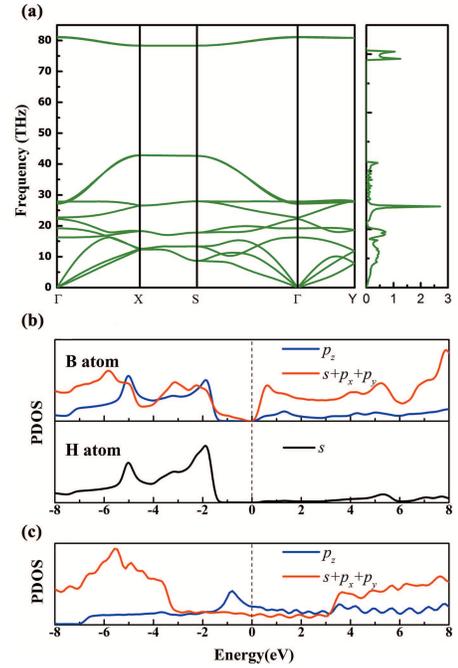}
\caption{\label{fig2}(a)Phonon band dispersions and density of states of the borophane. (b)PDOS of borophane. (c)PDOS of borophene.}
\end{figure}

Although borophene has been successfully synthesized on the Ag substrate, the free standing layer is not stable, meanwhile the unsaturated boron atom is expected to be oxidized and affected by the environments easily. There is an imaginary frequency in the phonon band dispersions of freestanding borophene, which means the borophene may exhibit instability against long-wavelength transversal vibrations\cite{mannix2015synthesis}. In order to examine the dynamic stability of borophane, the phonon band dispersions were calculated. As shown in Fig.~\ref{fig2}a, All imaginary frequencies have been removed after hydrogenation, which implies borophane is dynamic stable. Therefore, the hydrogenation is an effective way to improve the stability of the borophene.

To explain the stability of the borophane sheet, we consider the nature of their electronic bonding. Generally, in-plane bonds formed from overlapping \emph{sp$^2$} hybrids are stronger than out-of-plane $\pi$-bonds derived from \emph{p$_{z}$} orbitals, so a structure that optimally fills in-plane bonding states should be most preferable \cite{PhysRevLett.99.115501}. This conclusion has been confirmed in $\alpha$ and $\beta$ boron nanosheets, where the balance between two-centre and three-centre bonding assist to stabilize the structure.  Guided by this principle, we plotted the partial density of states (PDOS) of in-plane (\emph{s}, \emph{p$_{x}$} and \emph{p$_{y}$}) and out-plane (\emph{p$_{z}$}) states in Fig.~\ref{fig2}b. In borophene, each boron atom has six nearest neighbors, but has only three valance electrons, which is sufficiently localized to form strong covalent bonds but deficiently in numbers to crystalize the simple elemental structures \cite{PhysRevLett.105.215503}. We can see from PDOS of borophene, some of in-plane \emph{sp$^2$}antibonding states are occupied (due to the mixture of in-plane and out-plane states induced by buckling, the point between anitbonding and bonding state is not very clear, but it can be identified from flat triangle borophene, see Ref.  \cite{PhysRevLett.99.115501} ), it means that borophene has  a surplus of electrons in antibonding states and is prone to donate electrons,  which explain the reason of structural instability. When hydrogen is adsorbed on the boron atom, part of electrons is transferred to hydrogen atom from boron. From Bader analysis, it shows that 0.715e is transferred. The electron trandfer  on boron leads to two consequences: 1, the in-plane bonding states are completely filled, while the antibonding states are empty; 2, the out-plane bonding states are also fully occupied. As a result, the \emph{E$_{F}$} is exactly located at the zero point of the in-plane (as well as out-plane) PDOS as demonstrated in Fig.~\ref{fig2}c. Borophene is therefore structurally stabilized after hydrogenation.

\begin{figure}
\includegraphics{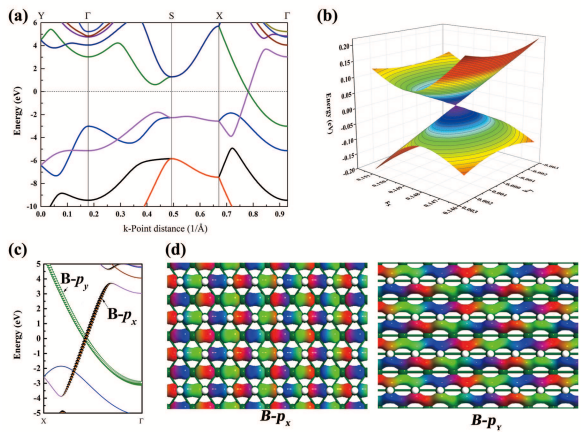}
\caption{\label{fig3}(a) the band structures and density of states of borophane. (b)Distorted Dirac cone formed in the vicinity of the Dirac point.(c)Orbitals of the linear dispersion relation along X and Y. Fermi level is set to zero. (d)Bloch states at the Dirac points of borophane.}
\end{figure}

We now turn our attentions to its electronic properties. As shown in Fig.~\ref{fig3}a, we can see that most of the states near Fermi level of borophene have been removed after hydrogenation, the remaining band states show prefect linear dispersion along $\Gamma$ to X with valence and conduction bands meeting at a single point around the Fermi level (zero states near the Fermi level, Fig.~\ref{fig2}b). In other words, borophane possesses a Dirac cone like graphene. The three-dimensional Dirac cone is shown in Fig.~\ref{fig3}b. Compared with most well-known graphene, the Dirac states in borophane exhibit couple of obvious differences and advantages. First of all, borophane exhibits two Dirac points in its first Brillouin zone rather than six in graphene, but only one of them is shown here due to the symmetry, which is located on the lines from $\Gamma$ to X. Secondly, this Dirac cone is distorted and asymmetric, with slopes of +72 eV{\AA} and -40 eV{\AA} around the Dirac cone. The Fermi velocity of the Dirac fermions was calculated by the expression $\upsilon_{F}=E(k)/(\hbar k)$. In the \emph{k$_{x}$} direction, the fermi velocity is 3.0$\times$10$^{6}$ m/s and -1.7$\times$10$^{6}$ m/s, it is surprised that they are 2-4 times higher than that of graphene (8.2$\times$10$^{5}$ m/s)\cite{trevisanutto2008ab,hwang2012fermi} or one order magnitude larger than some graphene allotrope \cite{wang2015phagraphene,malko2012competition}, and they are the record of the largest Fermi velocity until now. Such a ultra-high Fermi velocity is originated from unique bonding states, in-plane bonding (\emph{p$_{x,y}$}) nature (see following), rather than the out plane state \emph{p$_{z}$} like in graphene. Since the effective mass of carrier is calculated as $m^{*}=\hbar^{2}(\frac{d^{2}E(k)}{dk^{2}})^{-1}$
, the linear band dispersion near Fermi level leads to carriers massless in borophene. We thus expect that borophane also possesses ultra-high carrier mobility like, even over graphene.  Both ultrahigh fermi velocity and massless carrier character benefit future application of borophane.

In order to reveal the origin of the Dirac cone and its ultra-high Fermi velocity, the band dispersions and its Bloch states near the Dirac points have been analyzed based on the orbital symmetry. As shown in Fig.~\ref{fig3}c, two crossing bands at Fermi level are mainly contributed by the \emph{p$_{x}$} and \emph{p$_{y}$} orbitals of boron atoms, respectively. In the real space, the Bloch states at the Dirac points clearly confirm that in-plane \emph{p$_{x}$} and \emph{p$_{y}$} orbitals lead to the formation of a Dirac cone (Fig.~\ref{fig3}b). The bonds between \emph{p$_{x}$} orbitals forms $\sigma$-conjugated linear chain, while the coupling between \emph{p$_{y}$} orbitals leads to the formation of $\pi$ bonding like framework. This leads to remarkable anisotropy of Dirac fermion. The energy band crossings between \emph{p$_{x}$} and \emph{p$_{y}$} is responsible for the production of the Dirac cones. In particular, the Dirac cones of most of 2D hexagonal  materials mainly originate from the orbitals along the vertical direction of 2D plane, such as the \emph{p$_{z}$} orbital of graphene. In borophane, however, its Dirac cone is contributed from in-plane orbitals, no orbitals along $\textbf{z}$ direction. The in-plane state induced Dirac states is expected to be stable and robust under in-plane strain and vertical electric fields.

In-pane orbitals not only leads to exciting Dirac cones, but also are beneficial to the outstanding mechanical property. Due to the different orbital contribution along \textbf{x} and \textbf{y} direction and anisotropic structure buckling (wrinkle along \textbf{y} direction), the mechanical property are also anisotropic. The $\sigma$ bonds are the strongest type of covalent bonds, which forms the orbital framework along \textbf{x} direction near the Fermi level. Hence, the in-plane stiffness along \textbf{x} direction of borophane should be more powerful. Using the formula $Y=\frac{1}{S} \frac{\partial^{2}E}{\partial\varepsilon^{2}}$ (where S is the equilibrium area, E is total energy and $\varepsilon$ is the uniaxial applied strain), we can estimate the Young's modulus. The calculations show that the values along \textbf{x} and \textbf{y} directions are 200.2 and 129.03 GPa$\cdot$nm respectively, indicating anisotropic but outstanding mechanical feature, and also verified that the in-plane stiffness along \textbf{x} direction is indeed stronger (it is understandable since the wrinkle along \textbf{y} direction weakens the mechanical properties). It is noticed that the calculated  in-plane Young's modulus in borophane is lower than the reported values of borophene which is up to 398 GPa$\cdot$nm along the \textbf{x} direction\cite{mannix2015synthesis}. This is due to that the B-B bond length in borophane is significantly stretched after hydrogenation, see Figure 1. The Young's modulus of borophane, we can see, is lower than graphene (2.4 TPa) \cite{YM}, but comparable with steel (200 GPa).

\begin{figure}
\includegraphics{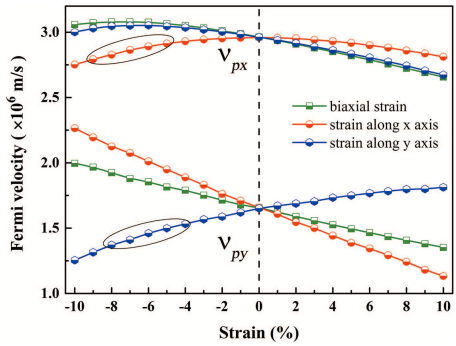}
\caption{\label{fig4}The response of the strain loading on the Fermi velocity of borophane. Two special lines are marked by the black circle.}
\end{figure}

In order to examine the stability and robustness of Dirac cone, external strain was applied on the borophane. Three types of strain have been considered: biaxial, uniaxial along the \textbf{x}- and \textbf{y}-directions. The strain was defined as $\varepsilon_{x}=(a-a_{0})/a_{0}$ and $\varepsilon_{x}=(b-b_{0})/b_{0}$, where $a_{0}$ and $b_{0}$ are the lattice constants of the strained and unstrained structures. For the uniaxial strain, the Poisson effect has been considered. It is found that the Dirac cone is robust regardless of strain types within -10\% (compression) to 10\% (expansion). However, the position and Fermi velocity are changed and exhibit high anisotropy. For the two Fermi velocities, since large one is originated from \emph{p$_{x}$}, lower one is from \emph{p$_{y}$} as shown in Fig.~\ref{fig3}, we thus name them as $\upsilon_{px}$ and $\upsilon_{py}$ respectively. Under biaxial strain, both velocities are linearly decreased with strain; namely tensile deformation decreases of the Fermi velocity, while compression leads to its increase. For uniaxial strain along the \textbf{x}-direction, the response to the Fermi velocities $\upsilon_{px}$ and $\upsilon_{py}$  is quite different. Both $\upsilon_{px}$ and $\upsilon_{py}$ are decreased under expansion loading, however $\upsilon_{px}$ is decreased while $\upsilon_{py}$ is increased under compression. Under the uniaxial strain along the \textbf{y}-direction, $\upsilon_{py}$ is linearly increased and $\upsilon_{py}$ is decreased. $\upsilon_{py}$ is more susceptible to strain while $\upsilon_{px}$ can be changed a little under the strain deformation.

The underlying mechanism of strain-tunable Fermi velocity can be contributed to the changes of orbital overlap. The general rule is that, the band overlap is increased (decreased) under compression (expansion) deformation. In Fig.~\ref{fig4}, two special lines go against this rule as indicated by the circle mark, namely $\upsilon_{py}$ under \textbf{y}-uniaxial strain and $\upsilon_{px}$ under \textbf{x}-uniaxial strain. When the strain loads along \textbf{x} axis, the slope of B-\emph{p$_{x}$} band is always depressed. The main reason is that the partial band near the high-symmetry points \textbf{X} and \textbf{$\Gamma$} is contributed by the B-\emph{p$_{z}$} and H-\emph{s} orbitals, the strain loading will change the proportion of B-\emph{p$_{x}$}, B-\emph{p$_{z}$} and H-\emph{s} orbitals in this band, and then affect the slopes of B-\emph{p$_{x}$} band. The other special line is the strain response of $\upsilon_{py}$ along the \textbf{y} axis. It is attributed to the arrangement of \emph{p$_{y}$} orbitals, which forms $\pi$-like bonds. During the tension process, these bonds are transformed into $\sigma$-like bonds, and then the overlap part of orbital will increase, which leads to the increase of the slops and Fermi velocities.

In summary, borophane is identified as a new 2D Dirac material based on the first-principles calculations. Compared with the borophene, this 2D material is dynamically stable induced by the electron transfer after hydrogenation. Its distorted Dirac cone is confirmed by the band structure and orbital symmetry. In particular, the Fermi velocities of borophane are calculated up to 3$\times$10$^{6}$ m/s, which is 2-4 times higher than that of graphene and is a record value until now. The Dirac cone is originated from the \emph{p$_{x}$} and \emph{p$_{y}$} orbitals of boron atoms, which is different from \emph{p$_{z}$} orbital Dirac cone in graphene, and responsible for ultra-high Fermi velocity and outstanding mechanical properties. The Dirac cone is robust agaist external strain, while the Fermi velocity is tunable. Borophane with Dirac fermions driven by in-plane orbitals broadens the family of Dirac material, and also can be a promising material for high-performance nanodevices.

\begin{acknowledgments}
We acknowledge generous grants of high-performance computer time from computing facility at Queensland University of Technology and Australian National Facility. L.X. gratefully acknowledge financial support by Natural Science Foundation for Young Scientists of Shanxi Province (2015021027) and the China Scholarship Council. A.D. greatly appreciates the Australian Research Council QEII Fellowship (DP110101239) and financial support of the Australian Research Council under Discovery Project (DP130102420). L.K. gratefully acknowledge financial support by the ARC Discovery Early Career Researcher Award (DE150101854)
\end{acknowledgments}

\nocite{*}
\bibliography{borophane}

\end{document}